\documentclass[letter,twocolumn]{jpsj2} 

\title{Increase in $d$-Wave Superconducting Transition Temperature near Imperfect Layer in Correlated Electron System}

\author{Tetsu Nobukuni, Hirono Kaneyasu\thanks{E-mail address: hirono@sci.u-hyogo.ac.jp}, Nobuyuki Shima, and Kenji Makoshi}

\inst{Graduate School of Material Science, University of Hyogo, 3-2-1 Koto, Kamigori, Ako, Hyogo 678-1297, Japan}

\abst{The effect of the site potential in an imperfect layer is studied in a $d$-wave layer superconductor on the basis of the electron correlation. The site potential binds electrons to the imperfect layer, and then, the superconductivity of the imperfect layer is independent of that of the bulk. We found that the superconducting transition temperature of the imperfect layer becomes higher than that of the bulk owing to the effect of the site potential. In this situation, the Fermi surface of the imperfect layer has a strong nesting feature leading to increases in both the antiferromagnetic spin fluctuation and the density of states near the Fermi level, which are in favor of $d$-wave pairing.}

\kword{$d$-wave superconductivity, interface, imperfect layer, defects, antiferromagnetic spin fluctuations}

\begin{document}
\maketitle
Recently, a curious phenomenon has been observed experimentally in the eutectic ruthenium oxide Sr$_2$RuO$_4$-Ru\cite{214Ru-1,214Ru-2,214Ru-3,214Ru-4} and the Ce compound CeIrIn$_5$\cite{115Ir}. In experiments\cite{214Ru-1,115Ir} on these two materials, the superconducting transition temperature $T_{\rm c, \rho}$ observed using zero resistance is higher than the bulk transition temperature $T_{\rm c, bulk}$ obtained by both the specific heat and the Meissner effect. In the $p$-wave superconductor Sr$_2$RuO$_4$-Ru, $T_{\rm c, \rho}$ is twice $T_{\rm c, bulk}=1.5$ K\cite{214Ru-1}. In the $d$-wave superconductor CeIrIn$_5$, $T_{\rm c, \rho}=1.2$ K is also higher than $T_{\rm c, bulk}=0.4$ K\cite{115Ir}.

The existence of Ru lamellae has been observed\cite{214Ru-1} in Sr$_2$RuO$_4$-Ru. Although $T_{\rm c, bulk}$ should be obtained by the superconducting transition of the entire system, $T_{\rm c, \rho}$ becomes higher than $T_{\rm c, bulk}$ in the case where the current is networked between interfaces of lamellae due to the interface superconducting state with a higher transition temperature. The interface superconducting state is bound to the interface and is separable from the bulk states. While the lamellar structures have not been observed in CeIrIn$_5$\cite{115Ir}, the two compounds have common properties as follows: One is that the two materials are unconventional superconductors with layer structures. The others are that both Ru $d$- and Ce $f$-electrons form the strongly correlated electron systems with cylindrical Fermi surfaces\cite{214Ru-FS1,214Ru-FS2,115Ir-FS}, and the mechanisms of the superconductivity in the bulk originate from the electron correlation with the 2-dimensional property.\cite{214Ru-Theory,115Ir-Theory}. Therefore, it is expected that the interface superconductivity also exists in CeIrIn$_5$, similarly to the case of Sr$_2$RuO$_4$-Ru. It is natural to theoretically study the mechanism of interface superconductivity with $T_{\rm c, \rho}$ higher than $T_{\rm c, bulk}$ on the basis of the effect of the electron correlation. 

In our previous study\cite{Kaneyasu}, we investigated the increase in the superconducting transition temperature of $d$-wave pairing at the surface on the basis of the effect of the electron correlation in the layer system. We found that the transition temperature at the surface becomes higher than $T_{\rm c, bulk}$ owing to the effect of the site potential at the surface atoms. The effect of the site potential increases both the antiferromagnetic spin fluctuation and the density of states at the surface, which are in favor of $d$-wave pairing. From these results, we suppose that the superconducting transition temperature becomes higher even at the interfaces of the imperfect layers. In an experiment on CeIrIn$_5$\cite{115Rh}, the effect of an impurity was reported as the mechanism behind $T_{\rm c, \rho}$ being higher than $T_{\rm c, bulk}$. Therefore, we propose the existence of imperfect layers formed by the impurities or defects in CeIrIn$_5$, and then, we investigate the effect of the site potential in an imperfect layer on the superconducting temperature of the $d$-wave pairing in this study. Hereafter, we call the superconducting state in the imperfect layer the imperfect-layer superconductivity. We apply the Green function theory\cite{Surface-Theory} of the surface superconductivity \cite{Kaneyasu} to the imperfect-layer superconductivity.

\begin{figure}
\begin{center}               
\includegraphics[height=8cm]{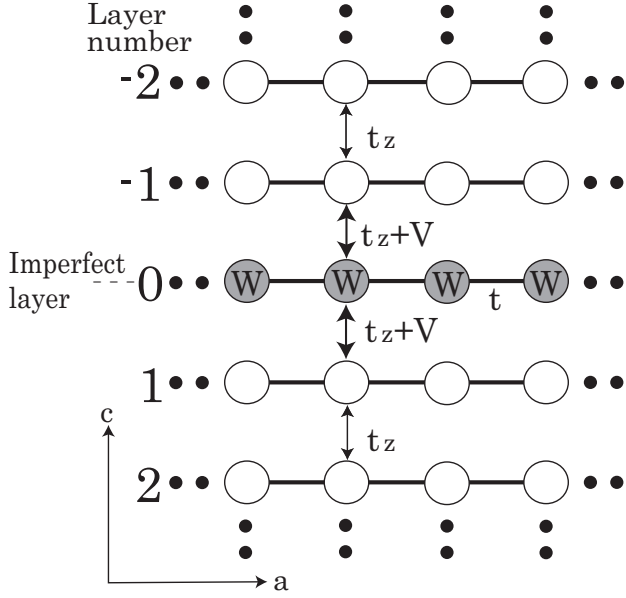} 
\end{center}
\caption{Multilayer system with an imperfect layer. The imperfect layer is positioned horizontally ($ab$ plane) in the bulk layers. The gray and white circles indicate atoms of the imperfect layer and those of the bulk layers, respectively. The nearest-neighbor transfer integral in the $ab$ plane is $t$. The nearest-neighbor transfer integral perpendicular to layers is $t_z$. The difference in transfer between the imperfect layer and the bulk layers is defined by $V$. The value of the site potential in the imperfect layer is denoted by $W$.}
\label{model}
\end{figure}

As shown in our model in Fig. \ref{model}, we consider a layer system with an imperfect layer, which is embedded parallel to the $ab$ plane. This imperfect layer is assumed to comprise atoms with the difference in site potential $W$. In CeIrIn$_5$, the electronic property mainly originates from the $f$-electrons\cite{115Ir,115Rh,115Co} of the Ce atoms occupying vertices of the tetragonal lattice. The cylindrical Fermi surface along the $k_z$-axis\cite{115Ir-FS} is denoted by the bulk dispersion:
\begin{eqnarray}
E(k_x, k_y, k_z)= 2t(\cos k_x + \cos k_y)+2t_{z} \cos k_{z} - \mu. 
\end{eqnarray}
Here, the nearest-neighbor transfer integral in the $ab$ plane is $t$. The nearest-neighbor transfer integral perpendicular to layers is $t_z$, which is much smaller than $t$ since the 2-dimensional property is strong. We set $t=1$ as the energy unit and $t_z=0.2$. We fix the chemical potential $\mu=-0.54$ at $T=0.0005$, and the value of $\mu$ gives an electron filling of 0.384 close to that of Ce-115 compounds.\cite{115Ir-FS,115Ir-Theory} The space symmetry is conserved in the $ab$ plane, although the translational symmetry is broken in the $z$-direction owing to the presence of the imperfect layer. We use the mixed Bloch-Wannier representation and Green function theory\cite{Surface-Theory}. Our representation is based on the Bloch representation in directions parallel to the imperfect layer. The number of layers is denoted by $m$, and the imperfect layer is numbered by $m=0$. Here, $t_z$ between the imperfect layer and the bulk layers at $m=\pm 1$ is reduced by the difference $V$ in transfer, which is fixed as $V=-0.25t_z$. The Green functions in this system including the imperfect layer are obtained using the Dyson-Gorkov equation:
\begin{eqnarray}
G_{m,n}^{I}(k_{\parallel},\omega_{l})=G_{m-n}^{B}(k_{\parallel},\omega_{l})+[G_{m}^{B}(k_{\parallel},\omega_{l})W ~~~~~~~~~~~~~~\nonumber \\ 
+ G_{m-1}^{B}(k_{\parallel},\omega_{l})V +G_{m+1}^{B}(k_{\parallel},\omega_{l})V]G_{0,n}^{I}(k_{\parallel},\omega_{l}) ~~~~~~~~~~~\nonumber \\
+G_{m}^{B}(k_{\parallel},\omega_{l})VG_{1,n}^{I}(k_{\parallel},\omega_{l})+G_{m}^{B}(k_{\parallel},\omega_{l})VG_{-1,n}^{I}(k_{\parallel},\omega_{l}),
\end{eqnarray}
where $\omega_{l}=\pi T(2l+1)$ is the fermion Matsubara frequency. The momentum in directions parallel to the imperfect layer, which corresponds to that in the $k_xk_y$ momentum space, is denoted by $k_{\parallel}$. The Green function in the system with the imperfect layer and that in the bulk system without the imperfect layer are defined as $G_m^B$ and $G_m^I$, respectively. We approximate the strong electron correlation by the on-site repulsion between f-electrons in the following Hubbard model: 
\begin{eqnarray}
H=\sum_{ i j  \sigma } t_{ij} c_{i\sigma }^{\dagger } c_{j \sigma} + U \sum_{i}n_{i \uparrow }n_{i \downarrow },
\end{eqnarray}
where $c_{i\sigma}^{\dagger}(c_{j \sigma})$ is the creation (annihilation) operator for the electron with the spin $\sigma$ at the site $i$ $(j)$. The transfer integral $t_{ij}$ denotes the nearest-neighbor hopping amplitude between the $i$-th and $j$-th sites. We set the value of the on-site Coulomb repulsion as $U=5.5$. We obtain the effective interaction for the $d$-wave pairing on the basis of the second-order perturbation theory with respect to $U$. 

In this study, we are interested in the imperfect-layer superconductivity, which is bound to the imperfect layer owing to the site potential $W$. The localized states, which are called imperfect-layer states, are energetically separable from the bulk states and appear under the condition that $|W|$ is larger than $\frac{7}{8}t_z$. This condition is obtained by the analysis of the Green function\cite{Surface-Theory}. When $W>\frac{7}{8}t_z$, the wave functions of superconductivity, as well as the Green function, are evanescent along the $z$-direction owing to the effect of $W$. Thus, the decay length of the Green function is sufficiently short along the $z$-direction in the energy range of predominant excitations of the superconductivity. This situation indicates that the superconducting equations can be separable between the imperfect layer and the bulk layers, and then, $T_{\rm c, imp}$ is also independent of $T_{\rm c, bulk}$. Therefore, the effective pairing interaction between layers can be neglected as an approximation, and the superconducting equations for each layer are obtained as
\begin{eqnarray}
\lambda_{m}\Delta_{m,m}(k_{\parallel},\omega_{l})=-\frac{T}{N_{\parallel}}\sum_{k_{\parallel}^{\prime},\omega^{\prime}_{l}}V_{m,m}(k_{\parallel}-k_{\parallel}^{\prime},\omega_{l}-\omega^{\prime}_{l})\nonumber \\G_{m,m}^{I}(k_{\parallel}^{\prime},\omega^{\prime}_{l}) 
G_{m,m}^{I}(-k_{\parallel}^{\prime},-\omega^{\prime}_{l})\Delta_{m,m}(k_{\parallel}^{\prime},\omega^{\prime}_{l}).~~~~~~~~~~~
\end{eqnarray}
Here, $\lambda_m$ indicates an eigenvalue in the $m$-th layer. The anomalous self-energy $\Delta_{m,m}$ has the $d_{\rm x^2-y^2}$-wave pairing symmetry because the $d_{\rm x^2-y^2}$-wave pairing is dominant in the bulk superconductivity of Ce-115 compounds. This result has been explained by experimental and theoretical studies\cite{115Ir-Theory, 115Ir-dx2-y2,115Co-dx2-y2,115Co-AF}. The effective interaction for the $d$-wave pairing in the $m$-th layer is given as 
\begin{eqnarray}
V_{m,m}(k_{\parallel}-k_{\parallel}^{\prime},\omega_{l}-\omega^{\prime}_{l})= 
U+U^2\chi_{m,m}(k_{\parallel}-k_{\parallel}^{\prime},\omega_{l}-\omega^{\prime}_{l}).
\end{eqnarray}
The susceptibilities $\chi_{m,n}$ are expressed as
\begin{eqnarray}
\chi_{m,n}(q_{\parallel},\Omega_{l} )= 
-\frac{T}{N_{\parallel}}\sum_{k_{\parallel},\omega_{l}}G_{m,n}^{I}(k_{\parallel}+q_{\parallel},\omega_{l}+\Omega_{l})G_{n,m}^{I}(k_{\parallel},\omega_{l}).
\end{eqnarray}
The diagonal component $\chi_{m,m}$ corresponds to the susceptibilities in the $m$-th layer. The boson Matsubara frequency is denoted by $\Omega_{l}=2\pi T l$ and the number of sites in the plane is defined as $N_{\parallel}$. The electron state in the $m$-th layer becomes the superconducting state at the transition temperature when the eigenvalue $\lambda_m$ of the $m$-th layer reaches 1. The behavior of $\lambda_m$ corresponds to the superconducting transition temperature, and thus, the large $\lambda_m$ corresponds to the high transition temperature. We calculate numerically the dependence of $\lambda_m$ on $m$ at $T=0.0005$. In our numerical calculations, we divide the first Brillouin zone of $k_xk_y$ into $128^2$ meshes and use $2048$ Matsubara frequencies.

\begin{figure}
\begin{center}                
\includegraphics[width=9cm]{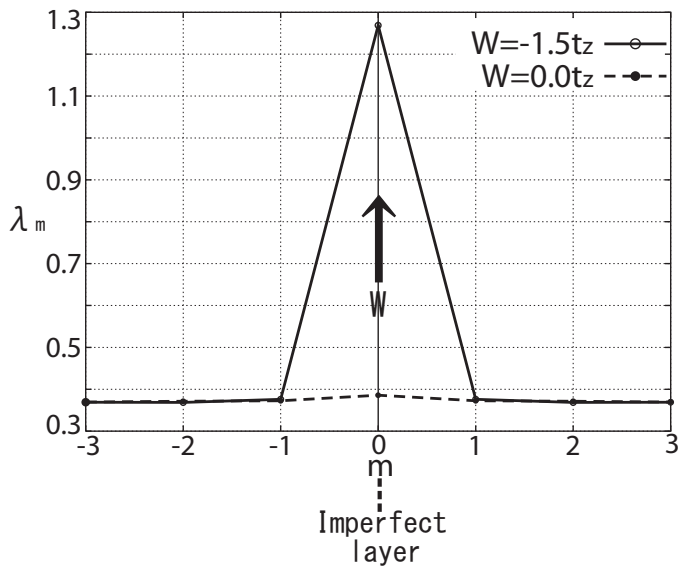} 
\caption{Dependence of $\lambda_m$ on $m$ at $T=0.0005$. The solid and dashed lines indicate $\lambda_m$ values at $W=-1.5t_z$ and $0$, respectively.}
\label{eigen-fig}
\end{center}
\end{figure}
In Fig. \ref{eigen-fig}, we show the dependence of $\lambda_m$ on $m$ at $T=0.0005$ as the numerical result. The eigenvalue $\lambda_0$ of the imperfect layer is markedly increased in the case that the value of the site potential is close to $W=-1.5t_z$, whereas $\lambda_m$ is not increased in the imperfect layer with $W=0$ and the other values except for $W=-1.5t_z$. The values of $\lambda_m$ of the other layers do not increase with $W$, and this value is almost equal to that of the bulk eigenvalue. This result indicates that $T_{\rm c, imp}$ becomes higher than $T_{\rm c, bulk}$ near $W=-1.5t_z$.

\begin{figure}
\begin{center}               
\includegraphics[width=8cm]{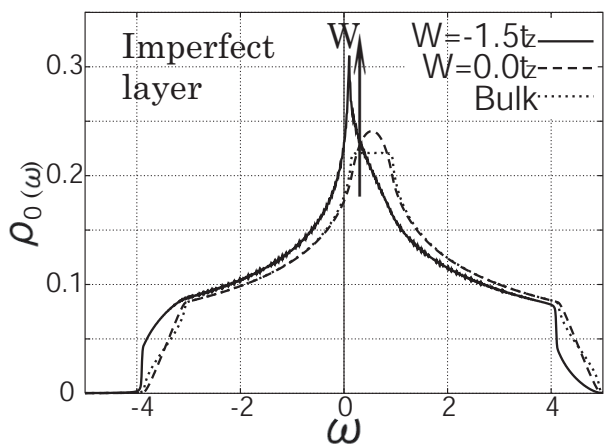}              
\caption{Effect of $W$ on $\rho_0(\omega )$ in the imperfect-layer state at $T=0.0005$.}
\label{dos-fig}
\end{center}
\end{figure}

To clarify the effect of $W$ on $T_{\rm c, imp}$, we show the density of states $\rho_{0}(\omega )$ in Fig. \ref{dos-fig}. The sharp peak of $\rho_{0}(\omega )$ locates near the Fermi level at $W=-1.5t_z$, and then, it induces $T_{\rm c, imp}$ to be higher than $T_{\rm  c, bulk}$. The sharp peak is not present near the Fermi level in the case of $W=0$ and is not close to the Fermi level at the other values except for $W=-1.5t_z$. As $-W$ becomes larger, the peak shifts to the lower energy side and grows sharply.

\begin{figure}
\begin{center}
\includegraphics[width=8cm]{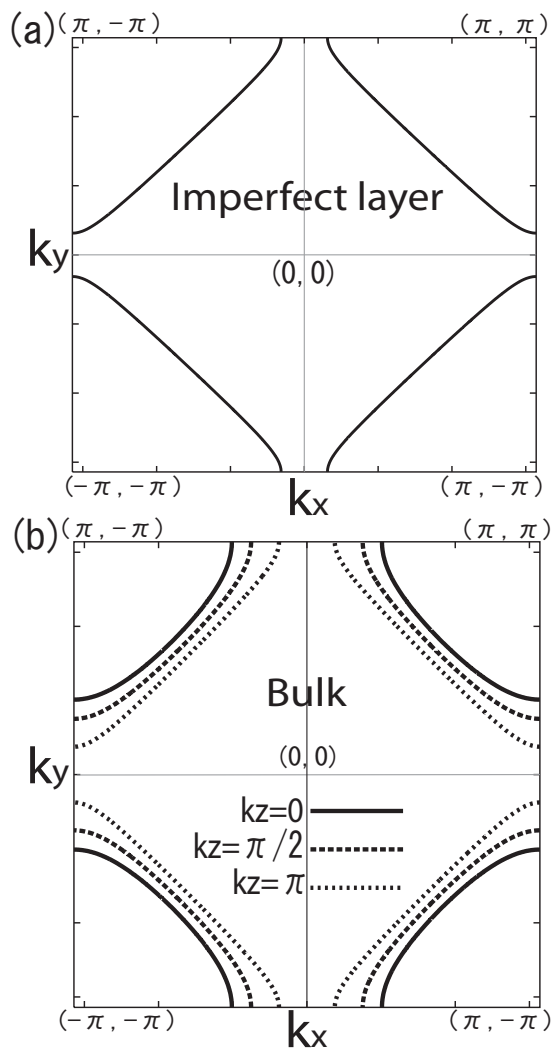}
\caption{(a) Fermi surface of the imperfect-layer state in the $k_xk_y$ momentum space at $T=0.0005$. It has the strong nesting feature. (b) Fermi surfaces in the $k_xk_y$ momentum space at $k_z=0$, $\pi/2$, and $\pi$ at $T=0.0005$ in the case of the bulk system without the imperfect layer.}  
\label{Fermi}
\end{center}
\end{figure}
\begin{figure}
\begin{center}
\includegraphics[width=9cm]{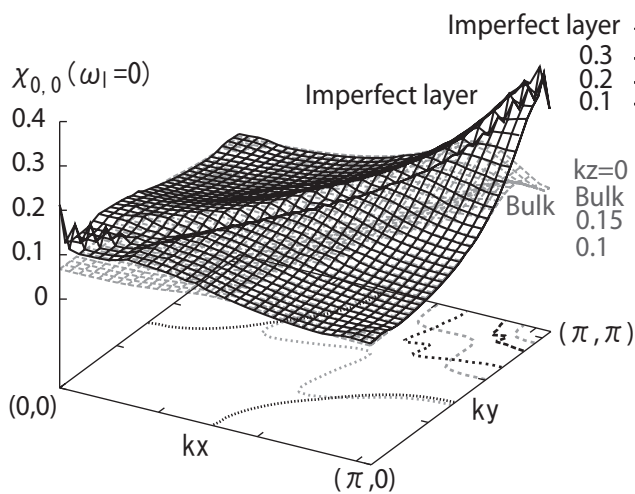}
\caption{Susceptibilities $\chi_{0, 0}(\omega_l=0)$ in the imperfect-layer state and the bulk state in the $k_xk_y$ momentum space at $T=0.0005$. Here, we show the susceptibility $\chi_{0, 0}(\omega_l=0)$ at $k_z=0$ in the bulk system without the imperfect layer. The antiferromagnetic spin fluctuations in the imperfect-layer state are enhanced at ($\pi, \pi$) owing to the effect of $W$.}
\label{Chi}
\end{center}
\end{figure}

We show the Fermi surface of the imperfect-layer state in the $k_xk_y$ momentum space in Fig. \ref{Fermi}(a). It has a $47.1\%$ area occupied by the electrons in the momentum space, and thus, the occupation of the Fermi surface corresponds to the near half-filling involving a strong nesting feature. The nesting feature induces the strong antiferromagnetic spin fluctuations at $(\pi, \pi)$, which appear in the susceptibility $\chi_{0, 0}(\omega_l=0)$ shown in Fig. \ref{Chi}. The momentum dependence of the antiferromagnetic spin fluctuations is in favor of the $d_{x^2-y^2}$-wave pairing. The above-mentioned features of both the strong antiferromagnetic spin fluctuations and the large $\rho_{0}(\omega)$ near the Fermi level markedly increase the  $\lambda_m$ of the imperfect layer, and then, the increase in $\lambda_m$ leads to that in $T_{\rm c, imp}$. The features of the imperfect-layer state are similar to the $d$-wave superconductivity in the high-$T_{\rm c}$ copper oxides\cite{high-Tc} near half-filling.

On the other hand, the features of the bulk states are different from that of the imperfect-layer state. We show the Fermi surfaces of the bulk state in the $k_xk_y$ momentum space at $k_z=0$, $\pi/2$, and $\pi$ in Fig. \ref{Fermi}(b). The electron filling in the bulk state is equal to 0.384. Although the Fermi surface at $k_z=\pi$ has the strong nesting feature similar to that of the imperfect-layer state, the nesting features are not so strong on the Fermi surfaces at the other values of $k_z$ including $k_z=0$ and $\pi/2$. While the strong nesting feature at $k_z=\pi$ produces the strong antiferromagnetic spin fluctuations leading to the high transition temperature, the Fermi surfaces at $k_z=0$ and $\pi/2$ do not. Therefore, $T_{\rm c, bulk}$ is lower than $T_{\rm c, imp}$ because it is determined by the effective interaction involving the entire set of Fermi surfaces at all values of $k_{\rm z}$.

In our conclusion, the effect of the site potential in the imperfect layer induces to be $T_{\rm c, imp}$ higher than $T_{\rm  c, bulk}$. The effect of the site potential produces the electron state bound to the imperfect layer, which is independent of the bulk states. The Fermi surface of the imperfect-layer state has the 2-dimensional nesting feature, which leads to increases in both the antiferromagnetic spin fluctuation and the density of states near the Fermi level, which are in favor of the $d_{x^2-y^2}$-wave superconductivity. We determined the value of the site potential leading to this situation and clarified that $T_{\rm c, imp}$ becomes higher than $T_{\rm  c, bulk}$ owing to the effect of the site potential in the imperfect layer. From the results, we propose the existence of an imperfect layer in CeIrIn$_5$, which gives $T_{\rm c, \rho}$ higher than $T_{\rm c, bulk}$.\cite{115Ir}

We are grateful to M. Sigrist for helpful discussions. The numerical calculations were carried out on SX8 at YITP, Kyoto University. This study was supported by Suzuki Foundation and Grant-in-Aid for Scientific Research Nos. 17069013 and 16101003 from the Ministry of Education, Culture, Sports, Science and Technology, Japan.

\end{document}